\documentclass[review]{siamart171218}

\usepackage{verbatim}
\usepackage{amsmath, amssymb, amsfonts, physics}
\usepackage{thmtools}
\usepackage{graphicx}
\usepackage{setspace}
\usepackage{color}
\usepackage{float}
\usepackage[utf8]{inputenc}
\usepackage[english]{babel}
\usepackage{framed}
\usepackage{amsmath,mathtools}
\usepackage{hyperref}

\newcommand{\dr}{\delta(r(t)-r)}

\newcommand{\dre}{\frac{\partial}{\partial r_E}}
\newcommand{\dri}{\frac{\partial}{\partial r_I}}

\allowdisplaybreaks
\nolinenumbers
\begin{document}

% ------------------------------------------------------------------------------
% Cover Page and ToC
% ------------------------------------------------------------------------------
\headers{Gaussian Equivalent Method}{Shoshana Chipman \& Brent Doiron}

\title{Dynamic Mean Field Theories for Nonlinear Noise in Recurrent Neuronal Networks \thanks{Funded by the Center for Living Systems}}

\author{
  Shoshana Chipman\thanks{University of Chicago, Department of Physics (\email{chipmansb@uchicago.edu}).}
  \and
  Brent Doiron\thanks{University of Chicago, Departments of Neurobiology and Statistics (\email{bdoiron@uchicago.edu}).}
}

\maketitle
\newpage

\tableofcontents
\newpage

% ------------------------------------------------------------------------------
\begin{abstract}
Strong, correlated noise in recurrent neural circuits often passes through nonlinear transfer functions, complicating dynamical mean-field analyses of complex phenomena such as transients and bifurcations. We introduce a method that replaces nonlinear functions of Ornstein–Uhlenbeck (OU) noise with a Gaussian-equivalent process matched in mean and covariance, and combine this with a lognormal moment closure for expansive nonlinearities to derive a closed dynamical mean-field theory for recurrent neuronal networks. The resulting theory captures order-one transients, fixed points, and shifts in bifurcation structure induced by noise, and outperforms standard linearization-based approximations in the strong-fluctuation regime. More broadly, the approach applies whenever dynamics depend smoothly on OU processes via nonlinear transformations, offering a tractable route to noise-dependent phase diagrams in computational neuroscience models. 
\end{abstract}
\section{Introduction}

The brain is a noisy dynamical system that exists far from equilibrium. Significant fluctuations are routine, and integral to functionality. One could even argue that a living brain has no steady state at all, and that its closest equivalent is a series of large-variability transients in response to stimuli. A rich tradition \cite{wilsoncowan, bardacidpaper, coombesbook, bressloffreview, coombesreview, cowan2011} exists of coarse-graining populations of neurons into mean-field models of neural firing rates. Much of this work, where it considers noise, does so as a finite-size effect, and as a small perturbation on the desired steady-state result. \cite{cowan2011} explicitly notes that this approximation is made in order to be able to use linear noise approximation as a ``main analytical tool''; others simply do not consider correlated input at the microscale. Noise fluctuations and variability are intrinsic to modeling brains. In this work, we derive a mean field theory capable of treating nonlinear noise dependence non-perturbatively, describing transient behavior in response to stimuli.\\

There are three distinct stages of averaging at play in neural mean field models: first, an average over time, to turn neuronal spike trains into firing rates; second, an average over space, to turn individual neural firing rates into a population average; third, an average over noise realizations, to produce a dynamic mean field theory for the evolution of population firing rates in terms of their statistical moments. During the first two stages of coarse-graining, recurrently coupled input -- the weighted sum of averaged firing rates --  is passed through some characteristic nonlinearity, called the transfer function. When the model includes inhibitory populations which can stabilize the system \cite{isn1, isn2}, the transfer function ought to be a power-law nonlinearity (experimental data, such as \cite{Priebe2004}, \cite{Miller2002NeuralNC}, and others indicate an exponent between 2.5 and 5); when there is no recourse to inhibition, some choice of sigmoid is traditionally used \cite{amari}. If the microscale model does not include correlated input, the macroscale model will have additive noise as a result of finite-size effects in the coarse-graining process \cite{bressloff2010}. If, however, input to the microscale is correlated, there will be noise inside of the transfer function at the macroscale, functioning as coarse-grained input to the firing rates. Moving the noise outside of the transfer function (\cite{tuckwell}) or assuming low noise intensity and solving for perturbations on the deterministic fixed point(\cite{Brent24}) constitute separate cases. Doing either reduces the SPDE to one of linear additive noise, instead of both additive and multiplicative noise with potential nonlinear dependence in both cases.
\\

 The systems in which we are interested routinely exhibit $\mathcal{O}(1)$ fluctuations in the firing rates, and so the simplicity of deriving mean field theories by perturbation comes at the price of discarding much of the intrinsic dynamics. Such methods \cite{sancho81,10.1063/1.525275} also require weak-noise approximations, or weak coupling and Gaussian firing rates, as in \cite{hennequin2016characterizingvariabilitynonlinearrecurrent}. Noise, treated inside the transfer function and on equal footing with firing rates, may not be weak. We seek a theory for $\mathcal{O}(1)$ fluctuations, without restrictions of noise, correlations, or couplings -- a treatment for nonlinear dependence on possibly-strong noise. We dub this the Gaussian equivalent method (GEM), and, in conjunction with a log-normal firing rate Ansatz, we construct a dynamic mean field theory (DMFT) for firing rates. This DMFT captures not only the fixed points, but also transients, of the underlying stochastic model, and inherits the bifurcation structure as well. We provide a prescription and estimation of accuracy of this method in treatments of generic nonlinear dependencies on Ornstein-Uhlenbeck (OU) noise. \\

 Section 2 derives the mesoscale model of population neural firing rates from the microscale model of instantaneous firing rates from the voltages. Section 3 derives the dynamic mean field theory for an E/I network with a thresholded quadratic nonlinearity, introduces the Gaussian equivalent method, and motivates lognormal moment closure for expansive nonlinearities. Section 4 discusses the predictive capacity of the DMFT on the effect of noise on the the bifurcations of the stochastic system. Section 5 considers a single neuronal population with a logistic transfer function. Section 6 presents analytic and numerical bounds for the goodness of the noise approximation for general nonlinearities. 
\section{Coarse-graining the voltage model}
%To demonstrate how shared fluctuations in a microscopic model necessitate putting noise inside the transfer function at the macroscale, 
We begin our analysis by coarse-graining a population of $N_E$ excitatory and $N_I$ inhibitory neurons into a macroscopic ensemble model. Consider the microscopic model of the population defined by:
\begin{equation}
\begin{aligned}
    r_\alpha^j &= \phi(V_\alpha^j),\\
    \tau_\alpha \frac{d}{dt} V_\alpha^j &= -V_\alpha^j + \sum_{k=1}^{N_E} W^{kj}_{\alpha E} r_E^k+ \sum_{k=1}^{N_I} W^{kj}_{\alpha I} r_I^k(t) +\mu_\alpha+ \gamma_\alpha \eta^j_\alpha(t).
\end{aligned}
\label{eq:micro}
\end{equation}
Here $r^j_\alpha$ is the firing rate of neuron $j$ in population $\alpha \in \{ E, I\}$, and $V_\alpha^j$ is the membrane potential of that neuron. Its dynamics are driven by the activity of other neurons $r^k_\beta$ in population $\beta \in \{E,I\}$, weighted by synaptic strength $W_{\alpha\beta}^{kj}$. The transfer function $\phi(V)$ maps the membrane potential to the firing rate. The stochastic process $\eta^j_\alpha(t)$ is a source of external fluctuations and $\mu_\alpha$ is a static input to neurons in population $\alpha$. Finally, $\gamma_\alpha$ and $\tau_\alpha$ set, respectively, the noise intensity and time scale of neurons in population $\alpha$.  \\

From our high-dimensional microscopic model we will derive a low-dimensional, aggregated, macroscopic one. The macroscopic model will consider state variables $r_E=\frac{1}{N_E}\sum_j r_E^j$ and $r_I=\frac{1}{N_I}\sum_j r_I^j$, the ensemble averaged firing rates over the $E$ and $I$ populations, respectively. We assume that neurons are all-to-all coupled, and that certain features (such as time scale and noise intensity) are identical across all neurons in a population. Further, we decompose the stochastic processes as:
\begin{equation*}
\eta_\alpha^j(t)=(1-\nu_\alpha)\chi^j_\alpha(t)+\nu_\alpha\eta_\alpha(t).
\end{equation*}
Here $\chi^j_\alpha(t)$ is a noise process that is unique to neuron $j$ in population $\alpha$, while the process $\eta_\alpha(t)$ is shared among all neurons in the population; the parameter $\nu_\alpha \in [0,1]$ sets the relative strength of these individual and shared processes. We take the shared process $\eta_\alpha(t)$ itself to decompose across the $E$ and $I$ populations as:
\begin{equation*}
\eta_\alpha(t)=(1-\rho)\chi_\alpha(t)+\rho \eta(t).
\end{equation*}
The $\chi_\alpha(t)$ are the fluctuations shared only within population $\alpha$. $\eta(t)$  are fluctuations seen by all populations. 
The motivation for considering shared fluctuations is to model low dimensional sources of external fluctuations that will correlate the joint activity of $\{r_\alpha^j\}$ in a low dimensional subspace, as is often reported in large-scale population recordings across a variety of brain regions \cite{ruff2020,urai2022,umakantha2021}.  

Under these assumptions, and in the limit of large $N_E$ and $N_I$, we find that the governing equations for the macroscopic variables $r_E$ and $r_I$ are (Appendix A):
\begin{equation}
\begin{aligned}
    \tau_E \frac{d}{dt} r_E &= - r_E + \phi\big (W_{EE} r_E + W_{EI} r_I+ \mu_E + \sigma_E \eta_E\big ),\\
    \tau_I \frac{d}{dt} r_I &= - r_I + \phi \big (W_{IE} r^E + W_{II} r_I+ \mu_I+ \sigma_I \eta_I \big),\\
    \tau_N \frac{d}{dt} \eta_\alpha &= -\eta_\alpha + \sqrt{\tau_N} \xi_\alpha(t),
\end{aligned}
\label{mysystem}
\end{equation}
where $\sigma_\alpha=\nu_\alpha\gamma_\alpha$ and $\langle \xi_\alpha(t)\xi_\beta(t')\rangle =[(1-\rho)\delta_{\alpha\beta}+\rho]\delta(t-t')$. %To begin we take $\phi(x) = \text{max}(0, x)^2$, modeling an expansive nonlinear response that is often reported \cite{}, though the techniques we develop will be generic (see Sec. \ref{logisticsec}). 
A key distinction between the microscopic system in Eq. \eqref{eq:micro} and the macroscopic system in Eq. \eqref{mysystem} is that the transfer function $\phi$ is applied to the inputs before temporal integration in the macroscopic system, in contrast to the microscopic system where $\phi$ is applied to their (linear) integration via the membrane potential ($V$) dynamics. 
We remark that our mean field approach necessitates the stochastic processes $\eta_\alpha$ being temporally correlated ($\tau_N >0$). This is so that $(r_E(t),r_I(t))$ is a well defined diffusive process, and is amenable to analysis. Physically, this implies that the stochastic inputs afferent to the network represent synaptic inputs that would naturally have some integration timescale ($\tau_N$). 

\begin{figure}[h]
    \centering
    \includegraphics[width=\linewidth]{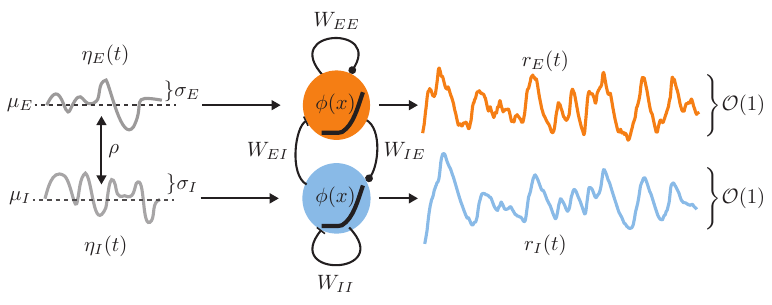}
    \caption{Realizations of the E/I network, and general schematic of the system. The parameters here are $\vec{W} = \{0.05, -0.75, 4, -3.5 \}$, $\mu_E= 5$, $\mu_I=9$, $\tau_E=6$, $\tau_I=20$, $\tau_N=1$, $\sigma_E=\sigma_I=1.0$, and correlation $\rho=0.5$}
    \label{fig:schematic}
\end{figure}

The central goal of our study is to build a theory for the statistics of $(r_E(t),r_I(t))$. If we take $\sigma_E$ and $\sigma_I$ to be sufficiently small, we could construct a mean field theory by linearizing around the deterministic fixed point, and considering the evolution of small perturbations away from the deterministic steady state. However, as we can see from Fig. \ref{fig:schematic}, the system in this regime has order 1 fluctuations, and so a meaningful amount of system behavior is captured in transients, and will be lost if we attempt to linearize around a fixed point. We also do not wish to assume weak noise. Even when averaging over realizations, $\mathcal{O}(1)$ transients persist in the mean. We therefore seek an alternative approximation to derive a mean field theory which can account for the effect of higher order moments in the mean, rather than solving for higher moments perturbatively through expansion.

%We also allow a time-varying input signal, $I_{\alpha}(t)$, which may be shared or not between the neuronal populations. 
%The $\phi$ of the firing rate equations may be re-scaled by a constant from the voltage transfer function, if it is thresholded. 

%We call the stochastic input $\eta_\alpha, \alpha \in \{E,I\}$, driven by a Wiener process $\xi_\alpha$. We take $\eta_\alpha$ to be white in time, and to have correlation $\rho$, which we understand as the normalized covariance
%\begin{equation*}
 %   \rho = \frac{\langle \eta_\alpha \eta_\beta\rangle - \langle \eta_\alpha \rangle \langle \eta_\beta\rangle}{\sqrt{\langle \eta^2_\alpha\rangle \langle \eta^2_\beta\rangle}}
%\end{equation*}

\section{Quadratic Transfer Function}
We turn our attention to the model and question which will occupy us, illustrated in Fig. \ref{fig:schematic}.

We make the approximation which avoids the thresholding of the piecewise nonlinearity: $\phi(x) = x^2$. This restricts our parameter space to  where the input to the transfer function is always non-negative. Expanding out our Langevin equations of Eq.\ref{mysystem}, our system becomes 
\begin{align*}
\centering
    \frac{d}{dt} r_E(t) &= f_E(r_E, r_I, t) + g_E(r_E, r_I, t)\eta_E(t) + k_E(r_E, r_I, t)\eta_E(t)^2 \\
    \frac{d}{dt} r_I(t) &= f_I(r_E, r_I, t) + g_I(r_E, r_I, t)\eta_I(t) + k_I(r_E, r_I, t)\eta_I(t)^2\\
    \tau_N \frac{d}{dt} \eta_E(t) &= -\eta_E(t) + \sqrt{\tau_N}\xi_E\\
    \tau_N \frac{d}{dt} \eta_I(t) &= -\eta_I(t) + \sqrt{\tau_N}\xi_I\\
   \left \langle \xi_\alpha(t) \xi_\beta(t') \right \rangle &= \delta(t-t')(\rho + (1-\rho)\delta_{\alpha \beta})\\
    \left \langle \eta_\alpha(t) \eta_\beta(t') \right \rangle &= e^{-|t-t'|/\tau_N}(\frac{1}{2}\delta_{\alpha \beta} + \frac{\rho}{2}(1-\delta_{\alpha \beta})) 
\end{align*}
Here, for notational ease, we have defined helper functions:
\begin{align*}
    f_\alpha(r_E, r_I, t) &= \frac{1}{\tau_\alpha}\bigl(-r_\alpha(t) + (W_{\alpha E}r_E(t) + W_{\alpha I}r_I(t) + \mu_\alpha + I_\alpha(t))^2\bigr)\\
    g_{\alpha}(r_E, r_I, t) &= \frac{2}{\tau_\alpha}\sigma_\alpha(W_{\alpha E}r_E(t) + W_{\alpha I}r_I(t) + \mu_\alpha + I_\alpha(t))\\
    k_{\alpha}(r_E, r_I, t) &= \frac{\sigma_\alpha^2}{\tau_\alpha}
\end{align*}
$I_\alpha(t)$ is a possible time-varying input signal to population $\alpha$. Our noise terms, the $\eta_\alpha(t)$, are OU colored-noise processes, driven by the Gaussian white noise (Wiener processes) $\xi_\alpha(t)$. The issue arises in the terms $\eta_\alpha(t)$ -- though well-defined as dynamical input fluctuations, the process of coarse-graining over them relies on Novikov's theorem, a critical support of which is Gaussianity of noise. Attempts to rectify this \cite{sancho81} by adding corrective terms to Novikov's theorem do so in expansions of noise intensity. This is precisely what we seek to avoid. Such a protocol may be infinitely precise at infinitely many terms of expansion, but requires truncation at some power of noise in practice. We will be satisfied to make an approximation which, while perhaps not infinitely precise, preserves noise correlations, and works for arbitrary nonlinearities. This is an alternative to linearization which does not require weak noise. In Section \ref{sec:DKL}, we demonstrate that this approximation is a reasonably good one for any non-Gaussian noise which is a smooth function of an Ornstein-Uhlenbeck process, though it may in principle work well for any generic noise. For the full derivation of the Fokker-Planck equation of this system, see Appendix \ref{app:manipulations}.
\subsection{Gaussian Equivalent Method}
We replace non-Gaussian $\eta_\alpha(t)^2$ with a Gaussian equivalent. Consider Gaussian $\gamma_\alpha(t)$, whose moments and correlations are entirely determined by those of $\eta_\alpha(t)^2$, and then define $\gamma^\ast_\alpha(t)$ as the zero-mean fluctuations around the mean of $\gamma_\alpha(t)$. The constraints on $\gamma_\alpha(t)$ are:
\begin{align*}
    \text{First moment: } \quad \left \langle \eta_\alpha(t)^2\right \rangle &= \bar{\gamma}_\alpha(t) = \frac{1}{2} \\ 
    \text{Second moment: }\quad \left \langle \eta_\alpha(t)^2 \eta_\beta(t')^2\right \rangle&=\left \langle \gamma_\alpha(t)\gamma_\beta(t')\right \rangle = e^{-2|t-t'|/\tau_N}(\frac{3}{4}\delta_{\alpha \beta} + (1-\delta_{\alpha \beta})).
\end{align*}
To meet these constraints, we propose that $\gamma_\alpha$ evolves according to
\begin{align*}
    \frac{d}{dt} \gamma_\alpha &= -\frac{2}{\tau_N}(\gamma_\alpha(t) - \frac{1}{2}) + \sqrt{\frac{2}{\tau_N}}\mathcal{\xi}_\alpha(t)\\
    \left \langle \tilde{\xi}_\alpha(t) \tilde{\xi}_\beta(t')\right \rangle &= \delta(t-t')(\delta_{\alpha \beta} + (1-\delta_{\alpha \beta}\rho^2)), 
\end{align*}
where $\gamma_\alpha(t)$ is driven by a new Wiener process $\tilde{\xi}_\alpha(t)$. 
 We will eventually take the white noise limit in the usual approximation while deriving the Fokker Planck equation. We define the white noise limit as: 
\begin{align}
    \left \langle \eta_\alpha(t) \eta_\beta(t')\right \rangle = X_{\alpha \beta}e^{-|t-t'|/\tau_N} \rightarrow X_{\alpha \beta} \delta\left(t-t'\right)
    \label{noisecorr}
\end{align}

It is prudent to take a moment to comment on this, as it is a subtle point. We understand $\eta_\alpha(t)$ as being drawn from a stationary probability distribution $P_\eta$ at all points in time. Consider the discrete-time equation:
\begin{align*}
    \eta_\alpha^{t+1} = \left(1-\frac{\Delta t}{\tau_N}\right) \eta_\alpha^t + \sqrt{\frac{\Delta t}{\tau_N}}\xi_\alpha^t
    \label{discreteeta}
\end{align*}
It becomes very clear that the ordering of limits matters, and the choice of limits gives rise to a choice of understandings of whiteness. If we take the limit of Eq. \ref{discreteeta} $\tau_N \rightarrow 0$ before $\Delta t \rightarrow 0$, we force ourselves into the Itô formalism, in which noise is inherently white. If we take $\Delta t \rightarrow 0$, we move back to the Stratonovich continuum limit, and whiten by taking $\tau_N \rightarrow 0$. We note also that the autocorrelation function of Eq. \ref{noisecorr} has an integral which is independent of $\tau_N$. The Stratonovich formalism preserves the $\tau_N$-independence of the integral of the autocorrelation, but destroys $P_\eta$, which transforms from a Gaussian to a Dirac delta function. In this work, we choose to take both limits simultaneously, $\frac{\Delta t}{\tau_N} \rightarrow\frac{0}{0}=1$. This destroys the $\tau_N$-independence of the integral of the autocorrelation, but preserves $P_\eta$. It should be noted that the ordering of limits is a free choice which must be made about what to preserve while whitening. Accordingly, we are working in neither the Itô nor Stratonovich interpretations, but in a kind of nonstandard Stratonovich. If one wished to move to the Itô formalism from this formalism, the drift term would not be the usual Stratonovich drift term, but would depend on specific model parameters. For the form of this drift term and further discussion of this unnamed formalism see \cite{weirdwhitenoise}.\\

Using the Fokker-Planck equation (see Appendix \ref{app:manipulations}) and integrating by parts, we find the DMFT for the first and second moments:
\begin{align}
    \partial_t \left \langle r_E(t) \right \rangle &= \left \langle h_E \right \rangle  + \frac{X_{EE}}{2}\left \langle g_E\frac{\partial g_E}{\partial r_E}\right \rangle + \frac{X_{EI}}{2}\left \langle g_I\frac{\partial g_E}{\partial r_I}\right \rangle \label{GEMr}\\
    \partial_t \left \langle r_I(t) \right \rangle &= \left \langle h_I \right \rangle  + \frac{X_{II}}{2}\left \langle g_I\frac{\partial g_I}{\partial r_I}\right \rangle + \frac{X_{EI}}{2}\left \langle g_E\frac{\partial g_I}{\partial r_E}\right \rangle\\    
    \partial_t \left \langle r_E(t)^2\right \rangle &= 2 \left \langle h_E r_E\right \rangle + X_{EE} \left \langle \frac{\partial g_E}{\partial r_E}g_E r_E\right \rangle + X_{EI} \left \langle \frac{\partial g_E}{\partial r_I}g_I r_E\right \rangle + X_{EE} \left \langle g_E^2\right \rangle + E_{EE}\left \langle k_E^2\right \rangle\\
    \partial_t \left \langle r_I(t)^2\right \rangle &= 2 \left \langle h_I r_I\right \rangle + X_{II} \left \langle \frac{\partial g_I}{\partial r_I}g_I r_I\right \rangle + X_{EI} \left \langle \frac{\partial g_I}{\partial r_E}g_E r_I\right \rangle + X_{II} \left \langle g_I^2\right \rangle + E_{II}\left \langle k_I^2\right \rangle\\
    \partial_t\left \langle r_E(t)r_I(t)\right \rangle &= \left \langle h_E r_I\right \rangle + \left \langle h_I r_E\right \rangle + \frac{X_{EE}}{2}\left \langle g_Er_I \frac{\partial g_E}{\partial r_E}\right \rangle + \frac{X_{II}}{2}\left \langle g_Ir_E \frac{\partial g_I}{\partial r_I}\right \rangle \\ &\quad \quad + \frac{X_{EI}}{2}\left \langle g_Ir_I \frac{\partial g_E}{\partial r_I}\right \rangle + \frac{X_{EI}}{2}\left \langle g_Er_E \frac{\partial g_I}{\partial r_E}\right \rangle  + X_{EI} \left \langle g_Eg_I\right \rangle + E_{EI}\left \langle k_Ek_I\right \rangle
\end{align}
This is the most compact way of writing out the system, but it is illuminating to ``unpack'' the helper functions, which hide nonlinearities. For example, the first excitatory moment given in Eq. \ref{GEMr}
\begin{align*}
    \frac{d}{dt} \left \langle r_E\right \rangle &= \left( \frac{2 W_{EE}\mu_E -1}{\tau_E} + \frac{2 X_{EE} \sigma_E^2 W_{EE}^2}{\tau_E^2} + \frac{2X_{EI} \sigma_E \sigma_I W_{IE} W_{EI}}{\tau_E\tau_I} \right)\left \langle r_E\right \rangle \\ 
    & \quad + \left(\frac{2 W_{EI}\mu_E}{\tau_E} + \frac{2 X_{EE} \sigma_E^2 W_{EE}W_{EI}}{\tau_E^2} + \frac{2 X_{EI} \sigma_E \sigma_I W_{EI}W_{II}}{\tau_E \tau_I} \right)\left \langle r_I\right \rangle \\
    & \quad + \frac{W_{EE}^2}{\tau_E}\left \langle r_E^2\right \rangle + \frac{W_{EI}^2}{\tau_E}\left \langle r_I^2\right \rangle + \frac{2W_{EE} W_{EI}}{\tau_E}\left \langle r_E r_I\right \rangle\\
    & \quad + \frac{\mu_E^2}{\tau_E} + \frac{2 W_{EE} X_{EE} \sigma_E^2 \mu_E}{\tau_E^2} + \frac{2 W_{EI}X_{EI} \sigma_E \sigma_I \mu_I}{\tau_E \tau_I} + \frac{\bar{\gamma}_E}{\tau_E}
\end{align*}
In the case of the quadratic nonlinearity, the $Nth$ moment will depend on the $(N+1)-th$ moment; thus, the evolution of the second moments of the firing rates depend on the third moments. 
\subsection{Moment Closure}
\begin{figure}[h!]
    \centering
    \includegraphics[width=\linewidth]{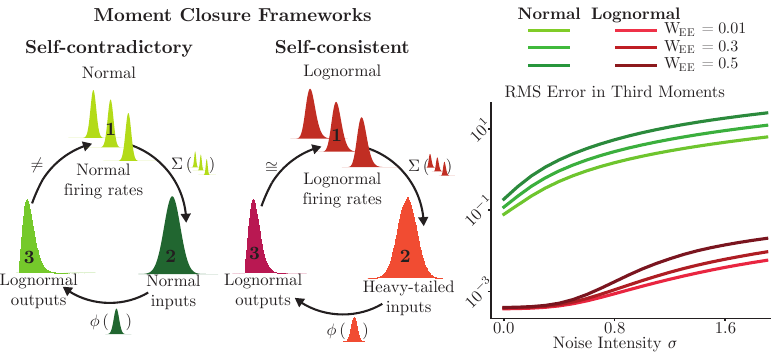}
    \caption{Left, a visualization of moment closure schemes for systems with expansive nonlinearities. Distributions (stage 1) are summed into inputs (stage 2) to an expansive nonlinearity, and pass through that nonlinearity into outputs (stage 3), which do or do not resemble the distributions in stage 1. Normal distributions sent through expansive nonlinearities develop tails; lognormal distributions had tails to begin with. Right: Root mean squared fractional error in all third moments, under normal vs lognormal moment closure.}
    \label{fig:moment closure}
\end{figure}

We are left with a mean field theory which must be closed. The linearization technique implicitly makes Gaussian assumptions about perturbations away from the mean; we could do the same here with the firing rates and close with Isserlis' theorem. However, this is unwise for two reasons: we do not consider noise to be a perturbation, and there is no a priori reason to assume Gaussian firing rates. In fact, assuming Gaussian firing rates would lead to a self-contradictory theory, seen on the left side of Fig. \ref{fig:moment closure}. Accordingly, we make the Ansatz that firing rates are jointly log-normally distributed, and will be gratified when this significantly reduces the error in the estimation of third moments (see Fig. \ref{fig:moment closure}, right side). Though weaker noise improves accuracy in both cases, the increase in error due to stronger connectivity is dwarfed by the error due to making the wrong moment closure Ansatz. An advantage of log-normality (though not a unique one) is that all moments of the log-normal distribution may be written in terms in the first two moments, and that all these generic moments are defined so long as the first two moments are nonzero. Biological plausibility demands that firing rates be positive, so this will not constrain our model.  We write the third moments of the firing rates as 
\begin{align*}
    \left \langle r_E^3 \right \rangle &= \left(\frac{\left \langle r_E^2 \right \rangle}{\left \langle r_E\right \rangle}\right)^3\\
    \left \langle r_I^3 \right \rangle &= \left(\frac{\left \langle r_I^2 \right \rangle}{\left \langle r_I\right \rangle}\right)^3\\
    \left \langle r_E^2 r_I \right \rangle &= \frac{\left\langle r_E^2\right \rangle \left \langle r_E r_I \right \rangle}{\left \langle r_E\right \rangle^2 \left \langle r_I\right \rangle}\\ 
    \left \langle r_E r_I^2 \right \rangle &= \frac{\left\langle r_I^2\right \rangle \left \langle r_E r_I \right \rangle}{\left \langle r_I\right \rangle^2 \left \langle r_E\right \rangle}\\ 
\end{align*}
This is purely due to properties of the log-normal distribution, and has no implicit small noise assumptions. A generic formula for closing arbitrary higher moments of lognormal distributions can be found in Appendix \ref{lognormal}.
\subsection{Comparison with the Linear Model}
The most obvious point of comparison for the GEM DMFT is Eq. \ref{mysystem} under weak-noise approximations. We are then able to perturbatively expand around a fixed point (in the manner of \cite{Brent24}) and consider the noise purely as additive. The linearized model is given by
\begin{align*}
    \tau_E \frac{d}{dt} \Delta r_E(t) &= - \Delta r_E(t) + L_E\left(W_{EE} \Delta r_E(t) + W_{EI}\Delta r_I(t)\right) + L_E \sigma_E \eta_E(t)\\
    \tau_I \frac{d}{dt} \Delta r_I(t) &= - \Delta r_I(t) + L_I\left(W_{IE} \Delta r_E(t) + W_{II}\Delta r_I(t)\right) + L_I \sigma_I \eta_I(t)\\
    L_\alpha &= \phi'\left(W_{\alpha E}r^*_E + W_{\alpha I}r^*_I + \mu_\alpha\right)\\
    \tau_N\frac{d\eta_\alpha}{dt} &= -\eta_\alpha +\sqrt{\tau_N}\xi_\alpha
\end{align*}
where $r^*_\alpha$ are the steady-state values, and $r_\alpha(t) = r^*_\alpha + \Delta r_\alpha(t)$. It is clear that $\frac{d}{dt} r_\alpha(t) = \frac{d}{dt} \Delta r_\alpha(t)$; we derive an evolutionary equation for the probability density.
\begin{align*}
f^\text{lin}_\alpha &= \frac{1}{\tau_\alpha}\left(-\Delta r_\alpha(t) + L_\alpha \sum_\beta W_{\alpha \beta} \Delta r_\beta(t) \right)\\
\partial_t P_\text{lin}(r_E, r_I, t) &= -\dre \left(f^\text{lin}_E P_\text{lin}\right) -\dri \left(f^\text{lin}_I P_\text{lin}\right) + \frac{X_{EE}}{2}\left(\frac{L_E \sigma_E}{\tau_E}\right)^2\frac{\partial^2 P_\text{lin}}{\partial r_E^2} \\ &\quad  \quad+ \frac{X_{II}}{2}\left(\frac{L_I \sigma_I}{\tau_I}\right)^2\frac{\partial^2 P_\text{lin}}{\partial r_I^2} + X_{EI}\frac{L_E L_I \sigma_E\sigma_I}{\tau_I \tau_E}\frac{\partial^2 P_\text{lin}}{\partial r_E \partial r_I}
\end{align*}
Using this, we can derive a mean field theory for all moments of the linear model:
\begin{align}
    \frac{d}{dt} \left \langle \Delta r_E \right \rangle &= \frac{1}{\tau_E}\left( L_E W_{EE}-1\right) \left \langle \Delta r_E\right \rangle + \frac{1}{\tau_E}L_E W_{EI}\left \langle \Delta r_I\right \rangle \label{linMFTr}\\
    \frac{d}{dt} \left \langle \Delta r_I \right \rangle &= \frac{1}{\tau_I}L_I W_{IE}\left \langle \Delta r_E\right \rangle + \frac{1}{\tau_I}\left( L_I W_{II}-1\right) \left \langle \Delta r_I\right \rangle \\
    \frac{d}{dt} \left \langle \Delta r_E^2\right \rangle &= \frac{1}{\tau_E}\left( L_E W_{EE}-1\right) \left \langle \Delta r_E^2\right \rangle + \frac{1}{\tau_E}L_E W_{EI}\left \langle \Delta( r_Er_I)\right \rangle + X_{EE}\left(\frac{L_E \sigma_E}{\tau_E}\right)^2\\
    \frac{d}{dt} \left \langle \Delta r_I^2 \right \rangle &= \frac{1}{\tau_I}L_I W_{IE}\left \langle \Delta (r_E r_I)\right \rangle + \frac{1}{\tau_I}\left( L_I W_{II}-1\right) \left \langle \Delta r_I^2\right \rangle + X_{II} \left(\frac{L_I \sigma_I}{\tau_I}\right)^2\\
    \frac{d}{dt} \left \langle \Delta (r_E r_I) \right \rangle &= \left(\frac{L_E W_{EE}}{\tau_E} + \frac{L_I W_{II}}{\tau_I}-\frac{\tau_E+\tau_I}{\tau_E \tau_I}\right) \left \langle \Delta (r_E r_I)\right \rangle + \frac{1}{\tau_E}L_E W_{EI}\left \langle \Delta r_I^2\right \rangle \\&\quad \quad + \frac{1}{\tau_I}L_I W_{IE}\left \langle \Delta r_E^2\right \rangle + X_{EI}\frac{L_E L_I \sigma_E\sigma_I}{\tau_I \tau_E}
\end{align}
 For ease of analytical comparison, we restrict to the unthresholded $\phi$ regime and consider unpacked eq. \ref{linMFTr}:
 \begin{align*}
    \frac{d}{dt} \left \langle \Delta r_E \right \rangle &= \frac{1}{\tau_E}\left(2W_{EE}^2 r^*_E + 2W_{EE}W_{EI}r^*_I + 2\mu_EW_{EE} -1\right) \left \langle \Delta r_E\right \rangle \\ & \quad \quad + \frac{2}{\tau_E}\left(W_{EE}W_{EI} r^*_E + W_{EI}^2r^*_I + \mu_EW_{EI}\right)\left \langle \Delta r_I\right \rangle 
 \end{align*}
And compare once more to unpacked eq. \ref{GEMr}.
\begin{align*}
    \frac{d}{dt} \left \langle r_E\right \rangle &= \left( \frac{2 W_{EE}\mu_E -1}{\tau_E} + \frac{2 X_{EE} \sigma_E^2 W_{EE}^2}{\tau_E^2} + \frac{2X_{EI} \sigma_E \sigma_I W_{IE} W_{EI}}{\tau_E\tau_I} \right)\left \langle r_E\right \rangle \\ 
    & \quad \quad + \left(\frac{2 W_{EI}\mu_E}{\tau_E} + \frac{2 X_{EE} \sigma_E^2 W_{EE}W_{EI}}{\tau_E^2} + \frac{2 X_{EI} \sigma_E \sigma_I W_{EI}W_{II}}{\tau_E \tau_I} \right)\left \langle r_I\right \rangle \\
    & \quad \quad + \frac{W_{EE}^2}{\tau_E}\left \langle r_E^2\right \rangle + \frac{W_{EI}^2}{\tau_E}\left \langle r_I^2\right \rangle + \frac{2W_{EE} W_{EI}}{\tau_E}\left \langle r_E r_I\right \rangle\\
    & \quad \quad + \frac{\mu_E^2}{\tau_E} + \frac{2 W_{EE} X_{EE} \sigma_E^2 \mu_E}{\tau_E^2} + \frac{2 W_{EI}X_{EI} \sigma_E \sigma_I \mu_I}{\tau_E \tau_I} + \frac{\bar{\gamma}_E}{\tau_N}
\end{align*}
These agree in the noiseless steady-state case, as we would expect. In the presence of noise, however, the disagreement is significant. The Gaussian equivalent mean field theory depends irreducibly on higher moments (as intended), which are solved simultaneously with the mean rather than perturbatively. In doing so, it also accounts for correlations in the noise -- the linear model only recognizes these (the cross terms, which go as $X_{EI}$ and $E_{EI}$) at higher moments of the theory, which do not contribute to the steady-state solution. The GEM theory also depends on noise intensity, and allows for correlation, which the linearization theory does not, and cannot do.\\

\begin{figure}[h!]
    \centering
\includegraphics[width=\linewidth]{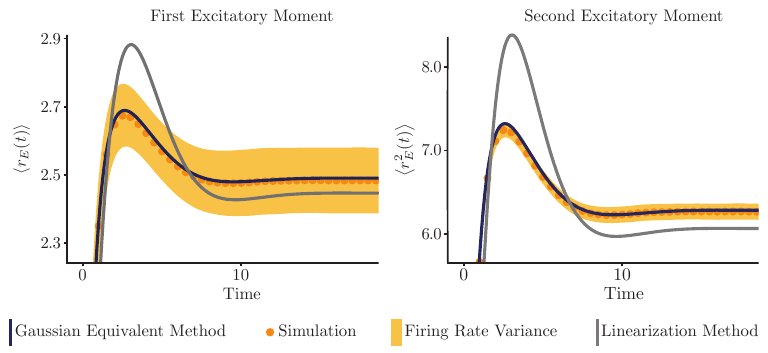}
     \caption{The Gaussian equivalent method and distribution technique successfully capture the underlying system, even when perturbative methods fail. The variance of the excitatory firing rate is shown on both plots, and the system is in a moderately high noise regime.   $\sigma_\alpha=0.7$.}
     \label{theorysimmatch}
\end{figure}

\begin{figure}
    \centering    \includegraphics[width=0.49\linewidth]{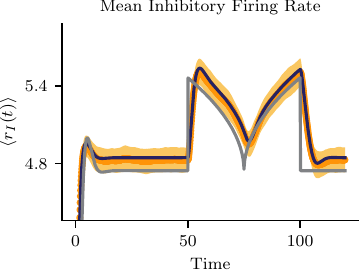}
    \includegraphics[width=0.5\linewidth]{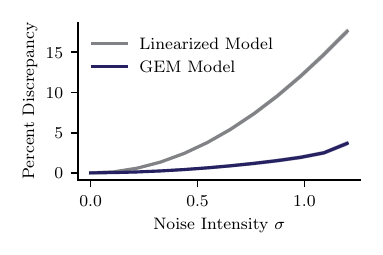}
    \caption{Left: the mean inhibitory firing rate of the system as it responds to a square-root pulse fed to the excitatory input. The GEM mean field theory captures fixed points and transients, including the counter-weight dip. Right: The root-mean-square percent error of the mean field theory, when compared to the underlying simulation, in five dimensions. Calculations do not include the transient, and consider the fixed point out to fifty time units. For each of the five dimensions of the mean field theory, percent error was calculated, and the result was the root sum-of-squares (distance in ``error space'').}
    \label{fig:discrepancy}
\end{figure}

To demonstrate the efficacy of the GEM DMFT compared with the linear model, we compare the predicted results of both of the theories to the simulations of the underlying system. Figure \ref{theorysimmatch} demonstrates that the Gaussian Equivalent Method model is more accurate at capturing the transients of underlying system. We reiterate that this failure is not surprising, since the transients are an $\mathcal{O}(1)$ fluctuation in the mean firing rate. 
   We can also compare the Gaussian equivalent method to the linear model by comparing the root-mean-square error in the steady-state as the noise intensity increases, as seen in Fig. \ref{fig:discrepancy}. We find that the GEM DMFT also outperforms in identifying the fixed points of the system  -- presumably, because it is able to account for the effect of noise in shifting the location of the fixed points, which the linear model cannot. The fluctuations are an integral part of the dynamics, and a model which captures them does better at capturing all parts of the system as a whole. 

\section{Bifurcation Analysis}
Wilson-Cowan models are known to support Hopf bifurcations. 
\begin{figure}
    \centering
    \includegraphics[width=\linewidth]{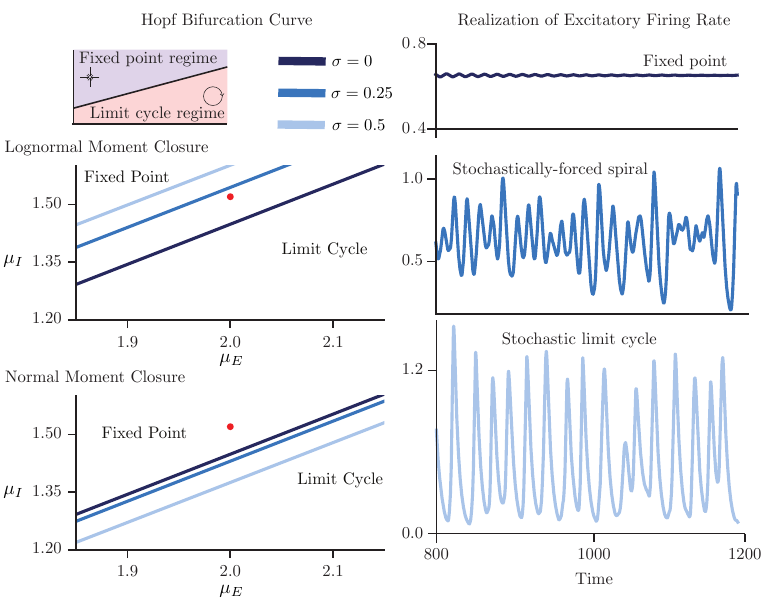}
    \caption{The location of the Hopf bifurcation as a function of inputs. $\vec{W}=\{2, -2.75, 4, -3.5\}$, $\mu_E=2$, $\mu_I=1.52$, $\tau_E=6$, $\tau_I=20$, $\tau_N=5$, $\rho=0.5$. Original bifurcation figures were generated in XPPAUTO \cite{AUTO}, and  regenerated nicely in PyPlot.}
    \label{fig:hopfbifurcationfigure}
\end{figure}
Hopf bifurcations in the presence of weak noise are a well-studied phenomenon 
\cite{Bashkirtseva2007StochasticBifurcationsRandom, Powanwe2021AmplitudePhaseDescription, Juel1997EffectNoisePitchfork, Triana2008HopfBifurcationsFluctuating}. When the noise is additive, one traditionally picks a sinusoidal rescaling of the key variables for the limit cycle regime. This then gives rise to a Fokker-Planck equation (for the new variables) with de facto multiplicative noise drift; by setting scaling parameters to leading order in noise, the asymptotic behavior and steady state distributions of the system can be ascertained. However, ``a general treatment of the fluctuations in the range of multiple macroscopic steady state has not been achieved with [this] approach'' \cite{Baras1982LimitCycle}. Noise makes bifurcations ``fuzzy'', giving rise to an ill-defined bifurcation region, instead of a bifurcation point. Multiplicative noise can create, shift, or destroy bifurcations altogether \cite{meunier1987noise}. The Hopf bifurcation of a Wilson Cowan system is, like all Hopf bifurcations, codimension-1, and accordingly ``stable''. Rather than destroying the bifurcation altogether (or giving rise to a totally new one), we anticipate the noise shifting the exact location of the limit cycle, giving rise to a bifurcation region rather than a point. The softness of the Hopf will add even further fuzziness, as a stochastic limit cycle that grows from radius zero can be difficult to distinguish from stochastic forcing away from a fixed point. We expect the incorporation of the noise intensity and higher moments into the mean field theory allows us to ``de-fuzz'' the impact of noise on the location of the bifurcation. Linearizing around a fixed point produces a mean field theory without any Hopf bifurcation at all.\\

As bifurcations are a qualitative change in system behavior, which the DMFT captures, we expect the bifurcations in the DMFT to reflect the bifurcations in the stochastic system, to some degree of accuracy. Figure \ref{fig:hopfbifurcationfigure} identifies the location of the Hopf bifurcation in the parameter space of the mean field theory, and compares the location of the bifurcation curve in the system with $\sigma_N=0$, (for which the theory is exact), $\sigma_N=0.25$, and $\sigma_N=0.5$. Moment closure is not unique, and often relies on intuition which is retroactively justified by efficacy. Some attempts \cite{kuehn2024preserving} have been made to put moment closure on a more rigorous framework for certain systems, but these are not generic. We leave the generic moment-closure problem for future work, but note that a good criteria of goodness suggested is that a moment closure regime should preserve the topological structure (namely, the bifurcations) of the underlying system. In the interest of testing our moment closure framework, we consider both the lognormal and normal moment closure. The former has noise push the onset of the limit cycle closer, at smaller $\mu_E$; the latter at higher $\mu_E$. We initialize the system near the noiseless bifurcation, in what should be the regime of the fixed point. The lognormal moment closure successfully predicts the early onset of the bifurcation.

\section{The Logistic Function}\label{logisticsec}
\begin{figure}
    \centering
    \includegraphics[width=\linewidth]{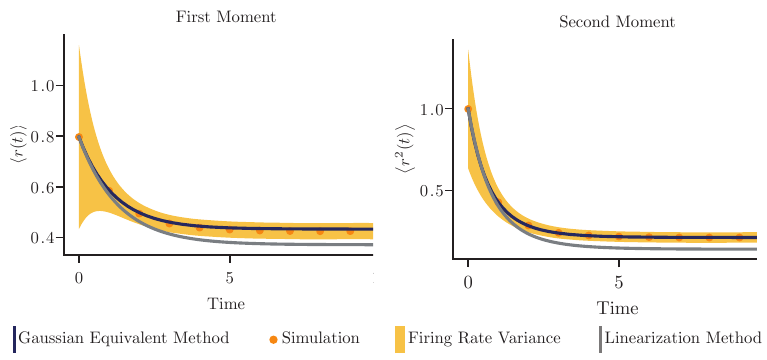}
    \caption{The mean firing rate (left) and variance (right) of the logistic transfer function system. We compare here the underlying simulated system, the mean field theory generated by linearization around a fixed point, and the mean field theory associated with the Gaussian equivalent method, up to third order in the Taylor expansion of the transfer function. The parameters  are $W=1, \tau=1, \tau_N=0.5, \sigma=1.75, \mu=-0.9$, variance is 0.032 and mean is 0.42}
    \label{logisticplots}
\end{figure}

The power of the Gaussian equivalent method is its generality; we wish to demonstrate that the piecewise quadratic is not a special case. In light of this, we briefly consider an alternate neural field model: that of a single excitatory population, passed through a sigmoidal transfer function. Such models abound in the literature \cite{bressloff2010, wilsoncowan, bardacidpaper, amari, beer1995, touboul2015noise}. For our transfer function $\phi$, we choose the logistic function.
\begin{equation}
\begin{aligned}
    \tau \frac{d}{dt} r &= - r(t) + \phi\left( W r(t) + \mu + \sigma \eta(t)\right)\\
    \tau_N \frac{d}{dt} \eta(t) &= -\eta(t) + \sqrt{\tau_N} \xi(t)\\
    \phi(x) &= \left( 1 + e^{-x}\right)^{-1}\\
    \left \langle \xi(t) \xi(t')\right \rangle &= \delta\left(t- t'\right)
\end{aligned}
\end{equation}
We could approximate the transfer function as linear, or linearize around the fixed point, but such processes rely on weak connectivity and weak noise. Instead, we choose to Taylor expand the transfer function.
\begin{equation}
    \tau \frac{d}{dt}r = -r + \frac{1}{2} + \frac{W r(t) + \mu + \sigma \eta(t)}{4} - \frac{\left(W r(t) + \mu + \sigma \eta(t)\right)^3}{48} + \mathcal{O}^5
    \label{logistic}
\end{equation}
We shall truncate at third order for the purposes of our analysis, and select parameters such that the fluctuations sample the nonlinearity up to cubic order, but not higher. This truncation is different than the perturbative noise treatment which we seek to avoid. In a perturbative treatment of noise, we assume that the noise is small relative to the state variables and couplings. Here, the noise may be arbitrarily large relative to those things -- but the entire input to the transfer function is small enough that the logistic function is well-sampled by the third order of its Taylor series. If this were not true, we would go to fifth, or higher order. We define new helper functions:
\begin{equation*}
    \begin{aligned}
        f(r) &= \frac{1}{\tau}\left(-r + \frac{1}{2} + \frac{W r + \mu}{4} -\frac{W^3 r^3 + 3W^2 r^2 \mu + 3 W r \mu^2 + \mu^3}{48}\right)\\
        g(r) &= \frac{\sigma}{16 \tau}\left(4 - W^2 r^2 - 2Wr\mu - \mu^2\right)\\
        k(r)&= -\frac{\sigma^2}{16\tau}\left(W r + \mu \right)\\
        j(r) &= -\frac{\sigma^3}{48\tau}
    \end{aligned}
\end{equation*}
Eq. \ref{logistic} becomes
\begin{equation*}
    \frac{d}{dt} r = f(r) + g(r)\eta(t) + k(r)\eta^2(t) + j(r) \eta^3(t)
    \label{newlogistic}
\end{equation*}
There are now two noise terms which must be replaced by new species of noise. While we need no longer contend with E/I correlations, there still exist correlations across noise species. By Isserlis' theorem \cite{Isserlis1918}, the first and third powers of noise have some correlation, and the replacement noise should maintain these relationships. This did not arise in the quadratic case, as the only cross-correlation among original and replacement noise would be equivalent to the third moment of Gaussian noise, which is zero. We may replace $\eta^2(t)$ with $\gamma(t)$, just as we did in the previous section, and similarly absorb the mean introduced by squaring a Gaussian random variable, $h(r)= f(r) + \left \langle \eta^2(t)\right \rangle k(r)$. We will also replace $\eta^3(t)$ with $\vartheta(t)$. This will be zero mean, as $\eta^3(t)$ is zero mean. However, there will now be cross-covariances, as $\left \langle \eta(t) \vartheta(t')\right \rangle = \left \langle \eta(t) \eta^3(t')\right \rangle \neq 0$. Consider the case of 
\begin{align*}
    \left \langle \eta(t) \eta(t')\right \rangle = \frac{D}{\tau_N}e^{-|t-t'|/\tau_N}
\end{align*}
In our model, $D=\tau_N/2$. The replacement noise should obey
\begin{equation}
\begin{aligned}
    \left \langle \vartheta(t) \vartheta(t') \right \rangle&=\left \langle \eta^3(t) \eta^3(t') \right \rangle = 6 \frac{D^3}{\tau_N^3}e^{-3|t-t'|/\tau_N} + 9 \frac{D^3}{\tau_N^3}e^{-|t-t'|/\tau_N} \rightarrow \frac{15}{8} \delta \left( t-t'\right) \\
    \left \langle \vartheta(t) \eta(t')\right \rangle &= \left \langle \eta^3(t) \eta(t')\right \rangle = 3 \frac{D^2}{\tau_N^2}e^{-|t-t'|/\tau_N} \rightarrow \frac{3}{4}\delta \left( t-t'\right)
\end{aligned}
\label{noise3corr}
\end{equation}
where we again take our probability-preserving white noise limit. The Langevin equation of $\vartheta(t)$ is:
\begin{align*}
    \vartheta(t) &= \upsilon(t) + \omega(t)\\
    \frac{d}{dt} \upsilon &= -\frac{3}{\tau_N} \upsilon(t) + \frac{6 D^{3/2}}{\tau_N^2}\xi_\upsilon\\
    \frac{d}{dt} \omega & = -\frac{1}{\tau_N}\omega(t) + \frac{6D}{\tau_N^2}\eta(t) + \left( \frac{18 D^3}{\tau_N^4} - \frac{1}{2}\left(\frac{6D}{\tau_N}\right)^2 \right)^{1/2} \xi_\omega,
\end{align*}
where we have denoted by subscript the independence of the white noise processes $\xi$. The mean field theory for the logistic transfer function neural field model is:
\begin{equation}
    \begin{aligned}
        \partial_t \left \langle r\right \rangle &= \left \langle h\right \rangle + \frac{1}{4}\left \langle g \frac{\partial g}{\partial r}\right \rangle + \frac{3}{8}\left \langle j \frac{\partial g}{\partial r}\right \rangle + \frac{1}{4}\left \langle k \frac{\partial k}{\partial r}\right \rangle \\
        \partial_t \left \langle r^2\right \rangle &= 2 \left \langle hr\right \rangle + \frac{1}{2}\left \langle g r \frac{\partial g}{\partial r}\right \rangle + \frac{1}{2}\left \langle g^2 \right \rangle + \frac{3}{4}\left \langle r j \frac{\partial g}{\partial r}\right \rangle + \frac{3}{2}\left \langle j g\right \rangle + \frac{1}{2}\left \langle kr  \frac{\partial k}{\partial r}\right \rangle\\&\quad \quad + \frac{1}{2}\left \langle k^2\right \rangle + \frac{15}{8}\left \langle j^2\right \rangle 
    \end{aligned}
    \label{logisticmft}
\end{equation}
Eq. \ref{logisticmft} contains third and fourth moments of the firing rate; we will once again be justified in using our log-normal moment closure Ansatz. As can be seen in Fig. \ref{logisticplots}, the Taylor expansion of the transfer function to third-order gives good agreement with the simulated system, even in the presence of large (order of $W$ or greater) noise. 
\section{Goodness of Approximation}\label{sec:DKL}
The Gaussian Equivalent Method (GEM) does fundamentally rely on the insistence that a distribution is well-approximated by a normal distribution which matches its first two moments. It would be useful for us to consider if this is true. In this work, we have modeled noise which is generated from Ornstein-Uhlenbeck processes and pushed through power laws with small exponents. We will now quantify how well a Gaussian models arbitrary powers of Gaussian random variables. Consider a zero-mean OU process, $\eta(t)$. Intuitively, the distribution of $\eta(t)^M$ should continue to have one peak, which is high relative to its tails. Odd $M$ should be better approximated by a Gaussian than even $M$.\\

We use the Kullback-Leibler divergence $D_{KL}$ to quantify how well one probability distribution is approximated by another. The $D_{KL}$ is given by
\begin{equation*}
    D_{KL}(P||Q) = \int P(x) \log\frac{P(x)}{Q(x)}dx
\end{equation*}
where $P$ is the approximated and $Q$ is the approximator. If $P$ and $Q$ are identical, $D_{KL}=0$. There is no upper bound, or absolute notion of goodness-of-agreement. The information loss ratio is the Kullback-Leibler divergence divided by the entropy of the unknown distribution $P$; since the $D_{KL}$ lacks a natural notion of goodness, we attempt to acquire one by comparing the amount of information lost by the $D_{KL}$  to the total amount of information in the original distribution. We approximate the entropy of the continuous distribution of $\eta^M$ by 
\begin{align*}
H_{P_\text{continuous}}=H_{P_\text{discrete}}-\log \Delta
\end{align*}
where $\Delta$ is the size of the bins in the numerical histogram \cite{Xie2012Lecture17}. Call $P$ the distribution of $\eta^M$. While the form of $P$ is not known  for generic $M$, all of its moments are \cite{winkelbauer2014momentsabsolutemomentsnormal}. Call $Q$ the Gaussian which matches the first two moments of $P$:
\begin{equation*}
    Q=\begin{cases}
          \mathcal{N}(s^M(M-1)!!, s^{2M}(2M-1)!!) \quad &\text{if} \, M \in 2\mathbb{Z} \\
          \mathcal{N}(0, s^{2M}(2M-1)!!)  \quad &\text{if} \, M \in 2\mathbb{Z} + 1
 \\
     \end{cases}
\end{equation*}
We can bound the loss ratio from below using Kullback's inequality \cite{wiki_kullbackineq}, which states that
\begin{align*}
    D_{KL}(P||Q) \geq \Psi^*_Q(\mu'_1(P)) \quad \forall M \in \mathbb{Z}
\end{align*}
$\mu'_1(P)$ is the first moment of $P$ (which we know to be 0 or $s^M(M-1)!!$), and $\Psi^*_Q$ is the convex-conjugate of the cumulant-generating function of $Q$. As entropy is always positive, it follows that 
\begin{align*}
    \frac{D_{KL}(P||Q)}{H(P)} \geq \frac{\Psi^*_Q(\mu'_1(P))}{H(P)} \quad \forall M \in \mathbb{Z}
\end{align*}
For ease, let us denote the first moment $\kappa_1$, and the second $\kappa_2$. Then the convex conjugate of the cumulant-generating functional of $Q$ is
\begin{align*}
    f^*(\kappa_1 t + \frac{\kappa_2 t^2}{2})=\frac{(y -\kappa_1)^2}{2 \kappa_2}
\end{align*}
which is zero at $y=\kappa_1$. The approximation may then be up to infinitely good, as $M$ has a minimum loss ratio of 0. \\

\begin{figure}
    \centering
    \includegraphics[width=0.5\linewidth]{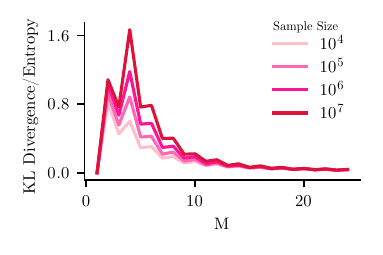}
    \caption{The loss ratio (Kullback-Leibler divergence divided by the entropy of the unknown distribution) for a variety of sample sizes -- a measurement of the information lost by approximation relative to the total amount of information contained in the original distribution. We pulled a sample of given size from a standard normal distribution, and raised it to the $M$. The mean and variance of the resulting set were used to generate a normal distribution. Samples drawn from this were compared to the underlying distribution. Sample size, and thus histogram bin size, has minimal impact on the result.}
    \label{KLD}
\end{figure}

For an upper bound, we turn to numerics, as shown in Fig. \ref{KLD}
Even powers of $M$ tend to do worse than odd ones, and the worst case seems to be where $M=4$ or so. We hypothesize that this is the tipping point, at which sharpness of the peak starts overpowering thickness of tails. The loss ratio asymptotes quickly to zero, implying that the GEM is quite good for large nonlinearities.
\section{Limits of the Technique}
As all approximations have their limitations, it is useful to consider the potential failures of the Gaussian Equivalent Method as an alternative to linearization.
\subsection{GEM is only as good as the moment closure Ansatz}
As the GEM is a technique designed to treat noise non-perturbatively, it is generally necessary to find some means of moment closure for higher moments in the mean field theory. In this work, firing rates are assumed to be jointly log-normally distributed. However, if this were \textit{not} true, the Ansatz would fail. In the event of a linear transfer function, for example, Gaussian input will become Gaussian output, and Isserlis' theorem could be used for moment closure. If the mean field theory entails the nonlinear dependence on a state variable which is not distributed according to a distribution whose higher moments are generically expressible in terms of its lower moments (such as the normal, lognormal, and exponential distributions), then the GEM will work, but is unlikely to produce an especially useful result, i.e., a mean field theory.

\subsection{Orders of Expansion}
\begin{figure}
    \centering
    \includegraphics[width=\linewidth]{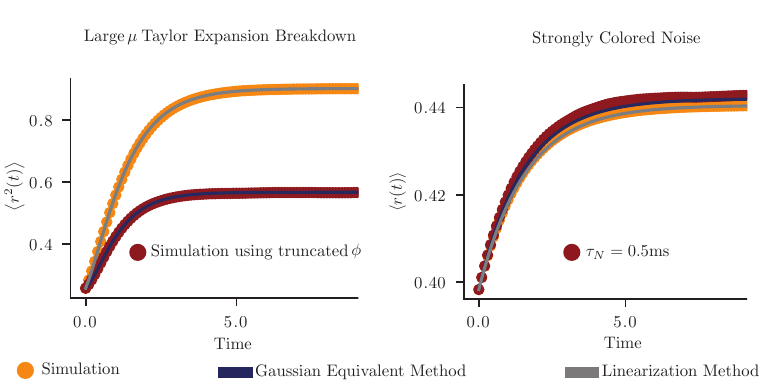}
    \caption{On the left, the second moment in a regime where the GEM fails due to an under-sampled transfer function. The parameters here are $W=1$, $\mu=2$, $\tau=1$, $\tau_N = 0.5$, $\sigma=0.1$. The GEM matches an alternative simulation where the transfer function is the third order Taylor expansion of the logistic function. On the right, the first moment of the firing rate in a regime where the noise too colored to approximate as white, and causes the mean field theory to fail. The parameters are $W=1.5$, $\mu=-0.9$, $\tau=1$, $\tau_N=50$, $\sigma=0.3$. The mean field theory matches a simulation identical but for a smaller $\tau_N$.}
    \label{logisticfailure}
\end{figure}

In section \ref{logisticsec}, we consider the case of noise inside of a logistic transfer function. We Taylor expand the function out to third order in its contents, and introduce a linear species of noise for each order of nonlinear noise. This leads to a mean field theory which describes the underlying system quite well, and significantly outperforms linearization when the noise is large. However, the goodness of the model is limited by the ability of the relevant-order truncation of the Taylor expansion to describe the underlying function. To illustrate this, in Fig.\ref{logisticfailure}, we increase the constant input, $\mu$, and find that the mean field theory is no longer an adequate description of the system. By plotting the the exact system of which the mean field theory is a description (the system whose transfer function is exactly the third-order Taylor expansion of the logistic function), we can see that the GEM mean field theory is an excellent match. Every additional order of the theory comes at the price of a new species of noise, and so this is a relevant practical limitation in implementation.
This extends more generally to other transfer functions. In some systems (such as the case of a polynomial of state variables an noise), it will be possible to exactly replace all nonlinearities of noise via the Gaussian equivalent method. But in many cases, some approximation of transfer function or restriction of parameter space may be necessary.

\subsection{Colored Noise}
In this paper, we work in a formalism wherein all noise is colored with temporal correlations, which we then proceed to ignore in the whitened limit. If the temporal correlation scale of our noise is too large, this limit will no longer be valid. Fully-colored noise breaks the Markovian assumptions which underpin the derivation of the Fokker-Planck formalism. To illustrate the effect of non-white noise, we consider Fig. \ref{logisticfailure}. Throughout this work, the noise has been normalized so that changing the timescale $\tau_N$ does not impact the noise intensity; for a selection of slower and slower noise, the Gaussian Equivalent Method captures the system less and less well. It should be noted, however, that this will also be an issue with linearization as an approximation scheme.

\subsection{Concessions to Smoothness May Obscure Phenomenology}
\begin{figure}
    \centering
    \includegraphics[width=\linewidth]{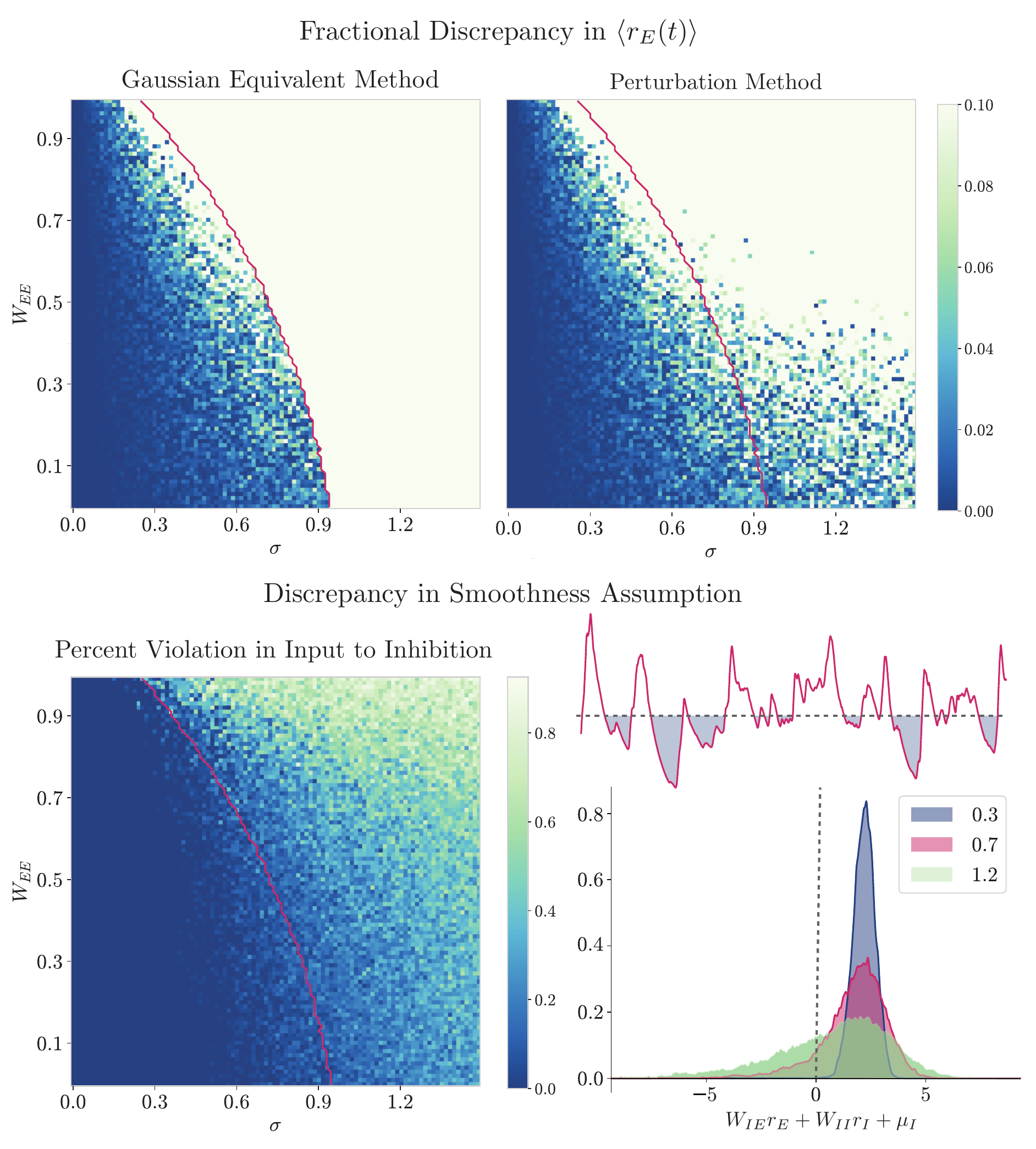}
    \caption{An examination of regions of validity of the GEM in the quadratic model: on the top, the discrepancy between $\text{var}(r_E)$ as found in the simulated equations versus the mean field theories, with the region of validity for the GEM overlaid in pink. On the bottom, the percent of a time series (post-transient) that the simulated input to $r_I$ is below zero.}
    \label{workability}
\end{figure}

Another related drawback of the GEM is that the nonlinearity in noise must be smooth. We do not present here a protocol for non-smooth functions. In the case of our underlying theory being thresholded (that is, the piecewise quadratic), we were obliged to work in a parameter regime that did not sample the thresholding. This constrained us into a particular regime, where the smoothness approximation was valid. For our model system, we restrict ourselves to a parameter regime where the full phenomenology of the system is not accessible. An active area of study in neural dynamics is that of quenching -- the ability of a population to suddenly go to zero firing rate. The Gaussian equivalent method cannot create a mean field theory of a piecewise-quadratic nonlinearity which describes this phenomenon. In order for the firing rates to go to zero, the system will have to live too close to the transition, and this is where the mean field theory diverges from the underlying case. At the very least, then, $\sum_\beta W_{\alpha \beta} r_\beta + \mu_\alpha \geq 0, \quad \alpha, \beta \in \{E, I\}$ -- though this condition is technically insufficient, as the addition of noise may send the expression below zero. If we consider Fig.\ref{workability}, we see that the GEM produces a mean field theory which diverges beyond this bound. While GEM outperforms linearization in regions of parameter space where the quadratic approximation is valid, it significantly underperforms outside of those regions. In the case of a transfer function approximation which does \emph{not} cause the mean field theory to become unstable when parameter assumptions are violated, care must be taken to ensure that work is done in the regime of validity.
\section{Conclusion}
We began with a principled model of neurons and sought to coarse-grain over space. Correlative structure in the noise coarse-grained to fluctuations in the population model for the firing rates, inside of transfer function. We further restricted the transfer function to be a piecewise power law, and forbade weak noise assumptions. This is a four-dimensional stochastic system driven by white noise. It could be solved as such, but in order to derive a mean field theory for the quantities of concern (firing rates), we would have to marginalize over the joint distribution of $r_\beta(t)$ and $\eta_\alpha(t)$. This joint distribution is not known, and would result in mean field equations with dependence on numerically-integrated constants, rather than clear system parameters able to provide intuition. We seek instead to treat  $\eta_\alpha(t)$ as colored noise which drives the system, and average with respect to it, in the manner of \cite{jordi}. This led to two things: the first is moment closure via the Ansatz of jointly-log-normal firing rates, and the other is the Gaussian Equivalent Method. The result is a DMFT which solves for all orders of moments in tandem, and is able to capture transient behavior as well as fixed points.\\

We further demonstrate that the GEM is robust for arbitrary integer powers of OU noise, and further consider the case of a logistic transfer function.  We work in a variant of the Stratonovich formalism which allows us to preserve the stationary density and non-temporal correlations of the noise in the white noise limit. Previous dismissals of this technique \cite{sancho81} do not account for the possibility of introducing multiple species of noise into the system of equations, and for our white noise limit. The Gaussian equivalent method should, in principle, work for arbitrary dependence of a system on arbitrary noise. It should be emphasized that at no point do we assume weakness of noise. The perturbative method solves for moments sequentially; the GEM solves for moments simultaneously.\\

We hope to soon extend this work into spatially extended systems, and to consider colored noise. We feel that it is a useful tool which allows for dynamical systems analyses of previously analytically-intractable systems. The ability of the GEM model to more accurately capture fixed points of the system in the presence of noise indicates the usefulness of the model in the consideration of noise-induced phase transitions in nonlinear systems. The location and stability of fixed points may shift in ways that are poorly captured by linearizing around the noiseless case. \\

The limitations of this technique on this class of systems are that it works only when nonlinearities are not piecewise. It also requires an accurate moment closure Ansatz, as it does not implicitly assume one.\\

\section*{Acknowledgments} We gratefully acknowledge support from the National Science Foundation through the Physics Frontier Center for Living Systems (PHY-2317138). We also thank Heather Cihak for fruitful discussion on Fourier transforms, and Satchal Postlewaite for his indubitable aesthetic flair.

\appendix
\section{Coarse-graining the microscopic  model}\label{app:coarsegrain}
Following \cite{bardbook} and \cite{bardacidpaper}, we write a set of implicit equations for the membrane potentials of neurons in the network:
\begin{align*}
    V^j_\alpha(t) = \sum_{\beta\in \{E,I\}} \int_{t_0}^t \frac{e^{-(t-s)/\tau_\alpha}}{\tau_\alpha}\left (\sum_kW^{jk}_{\alpha\beta} r_\beta^k(s) + \sigma_\alpha \eta_\alpha^j(s)\right )ds, \hspace{0.5cm} \alpha \in \{E,I\}.
\end{align*}
Here $W^{jk}_{\alpha\beta}$ is the connection strength from neuron $k$ in population $\beta$ to neuron $i$ in population $\alpha$. The firing rate $r_\beta^k(t)=\phi(V_\beta^k(t))$ for some nonlinear function $\phi$, so that there is an instantaneous mapping between membrane potential and firing rate dynamics.

In our study we assume that $W^{kj}_{\alpha\beta}$ depends only on the class of presynaptic and postsynaptic neurons (i.e the $\alpha$ and $\beta$) and we take the network to be all-to-all-coupled, with $W^{kj}_{\alpha\beta}= W_{\alpha\beta}/N$.  We can then write:
 \begin{comment}
\begin{align*}
    V^j_\alpha(t) = \sum_{\beta \in \{E, I\}}\sum_k \int_{t_0}^t \frac{e^{-(t-s)/\tau_\alpha}}{\tau_\alpha}(W_{\alpha \beta}r^k_\beta(s) + \sigma^j_\alpha \eta^j_\alpha(s))ds , \hspace{0.5cm} \alpha \in \{E,I\}.
\end{align*}
\end{comment}
\begin{align*}
    V^j_\alpha(t) = \sum_{\beta \in \{E, I\}} \int_{t_0}^t  \frac{e^{-(t-s)/\tau_\alpha}}{\tau_\alpha}(W_{\alpha \beta}r_\beta(s) + \sigma_\alpha \eta^j_\alpha(s))ds, \hspace{0.5cm} \alpha \in \{E,I\},
\end{align*}
where we have defined $r_\alpha=\frac{1}{N}\sum_k^N r^k_\alpha$ to be the population-average neural firing rate. The rate $r_\alpha$ is our macroscopic state variable that we will examine.

To derive the dynamical equation governing $r_\alpha$ we first consider $r_\alpha^j$:
\begin{align*}
    r^j_\alpha = \phi(V^j_\alpha) = \phi\left(\int_{t_0}^t  \frac{e^{-(t-s)/\tau_\alpha}}{\tau_\alpha}\sum_{\beta \in \{E, I\}} (W_{\alpha \beta}r_\beta(s) + \sigma_\alpha \eta^j_\alpha(s))ds\right), \hspace{0.25cm} \alpha \in \{E,I\}.
\end{align*}
We note that the stochastic process $\sigma_\alpha\eta^j_\alpha(s)$ is what in principle distinguishes the dynamics of $r^j_\alpha$ from $r^k_\beta$ (for $j \ne k$ and/or $\alpha \ne \beta$).
If we consider the infinitesimal change -- that is, $t_0 \approx t$, we can take $\tau_\alpha$ to be small and approximate:
\begin{align*}
    r^j_\alpha &\approx  \int_{t_0}^t \frac{e^{-(t-s)/\tau_\alpha}}{\tau_\alpha}  \phi\left(\sum_{\beta \in \{E, I\}} W_{\alpha \beta}r_\beta(s) + \sigma_\alpha \eta^j_\alpha(s)\right) ds, \hspace{0.25cm} \alpha \in \{E,I\}.
    \end{align*}
This allows us to write the population mean firing rate as:
\begin{align*}    
    r_\alpha &= \frac{1}{N_\alpha}\sum_j r^j_\alpha(t)\\ &= \frac{1}{N_\alpha} \sum_j \int_{t_0}^t  \frac{e^{-(t-s)/\tau_\alpha}}{\tau_\alpha}\phi\left(\sum_{\beta \in \{E, I\}} W_{\alpha \beta}r_\beta(s) + \sigma_\alpha \eta^j_\alpha(s)\right) ds, \hspace{0.25cm} \alpha \in \{E,I\}.
\end{align*}
We decompose the noise $\eta^j_\alpha(t)$ into a common factor model:
\begin{align*}
   \eta^j_\alpha(t) = \rho_\alpha\eta_\alpha(t) + \mu_\alpha + (1-\rho_\alpha)\chi^j_\alpha(t), 
\end{align*}
where the $\chi^j_\alpha(t)$ are a set of zero-mean Gaussian fluctuations of variance $s_\alpha^2$ that are private to neuron $j$ in population $\alpha$, and the $\eta_\alpha(t)$ are fluctuations common to all neuron belonging to population $\alpha$. %It is entirely possible to have additional noise which is seen by any sub-set of the $N$ neurons -- without loss of generality, however, we consider only the common signal and the individual signal. 
We further invoke \cite{bresslofftrick}, and switch the sum and the nonlinearity:
\begin{align*}    
    r_\alpha &=\int_{t_0}^t  \frac{e^{-(t-s)/\tau_\alpha}}{\tau_\alpha}\phi\left(\sum_{\beta \in \{E, I\}} W_{\alpha \beta}r_\beta(s) +  \sigma_\alpha\eta_\alpha(s) + \frac{1}{N}  \sum_j  \chi^j_\alpha(s)\right) ds, \hspace{0.25cm} \alpha \in \{E,I\}.
\end{align*}
The $\chi_\alpha^j$ are normally distributed random variables of zero mean; the sum over $j$ will therefore tend to zero as $N_\alpha \to \infty$. Only those fluctuations which are common to all members of the population survive, so that we have:
\begin{align*}    
    r_\alpha &=\int_{t_0}^t  \frac{e^{-(t-s)/\tau_\alpha}}{\tau_\alpha}\phi\left(\sum_{\beta \in \{E, I\}} w_{\alpha \beta}r_\beta(s) +  \sigma_\alpha\eta_\alpha(s)\right) ds, \hspace{0.25cm} \alpha \in \{E,I\}.
\end{align*}
In differential form this is given as:
\begin{align*}
    \tau_E \frac{d}{dt} r_E(t) &= - r_E + \phi(W_{EE} r_E(t) + W_{EI} r_I(t)+ \mu_E + \sigma_E \eta_E(t)),\\
    \tau_I \frac{d}{dt} r_I(t) &= - r_I + \phi(W_{IE} r_E(t) + W_{II} r_I(t)+\mu_I+ \sigma_I \eta_I(t)).\\
\end{align*}
Where we have separated out any constant component of the common input into $\mu_\alpha$
\section{Derivation of the Fokker-Planck Equation for the E/I Network}\label{app:manipulations}
We begin with the system of stochastic partial differential equations, as defined previously:
\begin{align*}
\centering
    \frac{d}{dt} r_E(t) &= f_E(r_E, r_I, t) + g_E(r_E, r_I, t)\eta_E(t) + k_E(r_E, r_I, t)\eta_E(t)^2 \\
    \frac{d}{dt} r_I(t) &= f_I(r_E, r_I, t) + g_I(r_E, r_I, t)\eta_I(t) + k_I(r_E, r_I, t)\eta_I(t)^2\\
    \tau_N \frac{d}{dt} \eta_E(t) &= -\eta_E(t) + \sqrt{\tau_N}\xi_E\\
    \tau_N \frac{d}{dt} \eta_I(t) &= -\eta_I(t) + \sqrt{\tau_N}\xi_I\\
   \left \langle \xi_\alpha(t) \xi_\beta(t') \right \rangle &= \delta(t-t')(\rho + (1-\rho)\delta_{\alpha \beta})\\
    \left \langle \eta_\alpha(t) \eta_\beta(t') \right \rangle &= e^{-|t-t'|/\tau_N}(\frac{1}{2}\delta_{\alpha \beta} + \frac{\rho}{2}(1-\delta_{\alpha \beta})) 
\end{align*}
Invoking the stochastic Liouville equation, we have that 
\begin{equation}
\begin{aligned}
    \partial_t P(r_E, r_I, t) &= \sum_{\alpha \in \{E, I\}}\frac{\partial}{\partial r_\alpha}\left \langle \dot{r}_\alpha \dr\right \rangle \\ & = -\frac{\partial}{\partial r_E}\left \langle \dot{r}_E \dr \right \rangle -\frac{\partial}{\partial r_I}\left \langle \dot{r}_I \dr \right \rangle 
\end{aligned}
\label{stochasticlouiville}
\end{equation}
where the triangular brackets in Eq. \ref{stochasticlouiville} are averages over initial conditions and all possible noise realizations. Van Kampen's lemma \cite{vankamp} states that for such brackets,
\begin{equation*}
    P(r_E, r_I, t) = \left \langle \dr \right \rangle.
\end{equation*}
We proceed, dropping explicit dependencies from our functions for clarity:
\begin{align}
    \partial_t P &=  -\dre \left \langle f_E \dr \right \rangle - \dri \left \langle f_I \dr \right \rangle \label{liou1}\\
    &\quad \quad - \dre \left \langle g_E\eta_E(t) \dr \right \rangle - \dri \left \langle g_I\eta_I(t) \dr \right \rangle \label{liou2} \\ 
    & \quad \quad  - \dre \left \langle k_E\eta_E(t)^2 \dr \right \rangle - \dri  \left \langle k_I\eta_I(t)^2 \dr \right \rangle \label{liou3}
\end{align}
Expression \ref{liou1} is the purely deterministic component of the Langevin equation, and straightforwardly renders:
\begin{align*}
    \left \langle f_\alpha \dr\right \rangle = f_\alpha P
\end{align*}
Expression \ref{liou2}, the first species of multiplicative noise, is somewhat more complicated, but we can continue to follow \cite{jordi}. We appeal to Novikov's lemma \cite{novikov}, which holds for zero-mean Gaussian noise such as $\eta_\alpha$, and find that 
\begin{equation}
\begin{aligned}
    \left \langle g_{\alpha } \eta_\alpha \dr \right \rangle &= \sum_\beta \int_0^t dt' \left \langle \eta_\alpha(t) \eta_\beta(t') \right \rangle \left \langle \frac{\delta g_{\alpha} \dr}{\delta \eta_\beta(t')} \right \rangle\\
    &= -\sum_{\beta \gamma}\frac{X_{\alpha \beta}}{2}\left \langle g_{\alpha} \frac{\partial}{\partial r_\gamma}\frac{\delta r_\gamma (t)}{\delta \eta_\beta(t')}\vert_{t=t'} \dr \right \rangle 
\end{aligned}
\label{leftoff}
\end{equation}
$\frac{\delta r_\alpha (t)}{\partial \eta_\beta(t')}\vert_{t=t'}$, the response function at equal times, is found by taking the variational derivative of the pseudo-integral:
\begin{equation*}
    r_\alpha(t) = \int_0^t f_\alpha(r_E, r_I, t') dt' +  \int_0^t g_{\alpha1}(r_E, r_I, t')\eta_\alpha(t')dt' +  \int_0^t g_{\alpha 2}(r_E, r_I, t') \eta_\alpha(t')^2 dt'
\end{equation*}
Here, we encounter an obvious problem: the response function at equal times will depend on a particular choice of noise realization. This will force the probability density of the firing rates to depend on choice of noise realization. Alternatively, we could treat $\eta_\alpha^2(t)$ as the averaging quantity, but this would break the necessary support for Novikov's theorem that random fluctuations be zero mean and Gaussian \cite{novikov}. The approximation of noise as an equivalent Gaussian is required at this step. \\
We may now define $\gamma^\ast_\alpha(t) = \gamma_\alpha(t) - \bar{\gamma}_\alpha$, and a new function, $h_\alpha(r_E, r_I, t) = f_\alpha(r_E, r_I, t) + \frac{\bar{\gamma}_\alpha}{\tau_\alpha}$, so that \ref{liou1}-\ref{liou3} become
\begin{align*}
    \partial_t P &=  -\dre \left \langle h_E \dr \right \rangle - \dri \left \langle h_I \dr \right \rangle \\
    &\quad \quad - \dre \left \langle g_E\eta_E(t) \dr \right \rangle - \dri \left \langle g_I\eta_I(t) \dr \right \rangle\\ 
    & \quad \quad  - \dre \left \langle k_E\gamma^\ast_E(t) \dr \right \rangle - \dri  \left \langle k_I\gamma^\ast_I(t)\dr \right \rangle
\end{align*}
We take the white noise limit of $\gamma_\alpha^\ast$:
\begin{align*}
    \left \langle \gamma^\ast_\alpha(t)\gamma^\ast_\beta(t')\right \rangle = E_{\alpha \beta} = X_{\alpha \alpha} \delta_{\alpha \beta} + (1-\delta_{\alpha \beta})2 X_{\alpha \alpha} X_{\beta \beta} \rho^2
\end{align*}
Returning to Eq. \ref{leftoff}, the response function at equal times is now clear:
\begin{equation*}
    \frac{\delta r_\alpha (t)}{\partial \eta_\beta(t')}\vert_{t=t'} = g_{\alpha}\delta_{\alpha \beta}
\end{equation*}
and thus,
\begin{align*}
    \left \langle g_{\alpha}\eta_\alpha(t) \dr\right \rangle &= -\frac{X_{\alpha \beta}}{2} g_{\alpha} \frac{\partial}{\partial r_\alpha}(g_{\alpha} P)\\
    \left \langle k_{\alpha}\eta^2_\alpha(t) \dr\right \rangle &= \left \langle k_{\alpha} \gamma^\ast_\alpha(t) \dr \right \rangle = - \frac{E_{\alpha \beta}}{2} k_{\alpha} \frac{\partial}{\partial r_\alpha}(k_{\alpha} P)
\end{align*}
From this, we have the Fokker-Planck equation for this system:
\begin{equation}
\begin{aligned}
    \partial_t P(r_E, r_I, t) &= -\dre(h_E P) + \frac{X_{EE}}{2}\dre g_E\dre(g_EP)+ \frac{X_{EI}}{2}\dre g_E\dri(g_IP) \\ & \quad \quad  -\dri(h_I P) + \frac{X_{II}}{2}\dri g_I\dri(g_IP)+ \frac{X_{EI}}{2}\dri g_I\dre (g_I P)\\ & \quad \quad
 + \frac{E_{EE}}{2}\dre k_E\dre(k_EP)+ \frac{E_{EI}}{2}\dre k_E\dri(k_IP) \\ & \quad \quad  + \frac{E_{II}}{2}\dri k_I\dri(k_IP)+ \frac{E_{EI}}{2}\dri k_I\dre (k_I P)
\end{aligned}
\end{equation}
\section{The \textit{M}th case}\label{app:generalcase}
Let us say that we have some OU-process, $\eta$,
\begin{align}
    \frac{d}{dt}\eta(t) = -\theta \eta(t) + A\xi(t).
    \label{xi}
\end{align}
where $\xi(t)$is some Wiener process drawn from the standard normal distribution, potentially with correlations. We present a recipe for replacing an arbitrary power of $\eta(t)$ using the Gaussian equivalent method.\\

We consider an OU process $\gamma(t)$. We wish $\gamma$ to match the first two moments and correlations of $\eta^M(t)$. The OU process $\eta(t)$ has the stationary probability distribution of a centered normal distribution, with variance $A^2/2\theta$. The $M$th central moment of the Gaussian is known, and is given by \cite{winkelbauer2014momentsabsolutemomentsnormal}:
\begin{equation*}
    \left \langle \mu^M \right \rangle=\begin{cases}
          s^M(M-1)!! \quad &\text{if} \, M \in 2\mathbb{Z} \\
          0 \quad &\text{if} \, M \in 2\mathbb{Z} + 1
 \\
     \end{cases}
\end{equation*}
\subsection{The correlations}
We seek to preserve the correlations of $\eta^M(t)$ in $\gamma(t)$. While in this work, we whiten $\eta$ for analytical ease, in principle there are temporal correlations which should be maintained. Consider the Fourier transform of Eq. \ref{xi}:
\begin{align}
    \hat{\eta}(\omega) = \frac{A \hat{\xi}}{\theta + i\omega} = \frac{1}{\sqrt{2 \pi}} \int e^{-i \omega t} \eta(t) dt
    \label{fourierxi}
\end{align}
We seek to match correlations; thus, $C^{\alpha \beta}(\tau) = \left \langle \gamma_\alpha(t) \gamma_\beta(t + \tau)\right \rangle = \left \langle \eta_\alpha^M(t) \eta_\beta^M(t+ \tau)\right \rangle$. Subscripts indicate species (such as excitation/inhibition). They may also indicate different species of replacement noise -- as seen in Sec. \ref{logisticsec}, if there are many noise terms to be replaced, their cross-correlations will have to be maintained. The correlation function is a convolution in time; therefore the Fourier transform of the correlation function is the product of the Fourier transforms of those functions being convolved. We denote the Fourier transform of the correlation function $S^{\alpha \beta}(\omega)$, the structure function.
\begin{align*}
    S^{\alpha \beta}(\omega) &= \hat{\gamma}_\alpha(\omega) \hat{\gamma}_\beta(\omega') = \mathcal{F}(\eta_\alpha^M(t))\mathcal{F}(\eta_\beta^M(t'))\\
    \hat{\gamma}_\alpha(\omega) &= \mathcal{F}(\eta_\alpha^M(t)) = \mathcal{F}(\prod^M\eta_\alpha(t))=\hat{\eta}_\alpha(\omega) \overset{M}{\ast} \hat{\eta}_\alpha(\omega)
\end{align*}
where we use $\overset{M}{\ast} $ to indicate $M$ convolutions. Using \ref{fourierxi}, we have 
\begin{align*}
    \hat{\gamma}_\alpha(\omega) =\int_{-\infty}^\infty \hat{\eta}(q_1)[\int_{-\infty}^\infty \hat{\eta}(q_2)[...[\int_{-\infty}^\infty \hat{\eta} (q_M)\hat{\eta}(\omega - \sum_i q_i)dq_M]...]dq_1
\end{align*}
Let us consider the $M$th convolution. Write $\omega_{M-1} = \omega -\sum_i^{M-1} q_i$. Then,
\begin{align*}
    \int_{-\infty}^\infty \hat{\eta}(q_M) \hat{\eta}(\omega_{M-1}-q_M) dq_M  &= A^2 \hat{\xi^2} \int_{-\infty}^\infty \frac{dq_M}{(\theta + i q_M)(\theta + i \omega_{M-1} - i q_M)} \\&= A^2 \hat{\xi^2} \frac{2 \pi}{2 \theta + i \omega_{M-2}-iq_{M-1}}
\end{align*}
Thus, we can say that 
\begin{align*}
    \hat{\gamma}_\alpha(\omega) &= \frac{A^M (2 \pi)^{M -1}\hat{\xi}^M}{M \theta + i \omega} 
\end{align*}
The inverse Fourier transform of the $M$th convolution of the Fourier transform of a standard normal random variable with some correlation will converge to a standard normal random variable, with the correlations associated with $\xi^M$ -- we shall call this new Wiener process $\tilde{\xi}$.
Thus, we know that the form of $\gamma(t)$ is given by
\begin{align*}
    \frac{d}{dt} \gamma(t) = -\theta (\gamma(t) - \Bar{\gamma}) + G \tilde{\xi}(t)
\end{align*}
that is, $\gamma(t)$ is an OU process, with some mean and variance. $\theta$ is uniquely determined by $M$ and $\eta(t)$. 
\subsection{The moments}
We first restrict the first moment of $\gamma(t)$, and declare that $\left \langle \gamma(t)\right \rangle = \left \langle \eta^M(t)\right \rangle$. Thus, 
\begin{equation*}
    \bar{\gamma}=\begin{cases}
          (\frac{A^2}{2\theta})^{M/2}(M-1)!! \quad &\text{if} \, M \in 2\mathbb{Z} \\
          0 \quad &\text{if} \, M \in 2\mathbb{Z} + 1
 \\
     \end{cases}
\end{equation*}
We may now restrict the second moment. 
\begin{align*}
    \left \langle \gamma^2 (t)\right \rangle = \frac{G^2}{2\theta} + \bar{\gamma}^2= (\frac{A^2}{2\theta})^{M}(2M-1)!!
\end{align*}
which binds the amplitude $G$. Thus, we can say
\begin{equation*}
    \frac{d}{dt} \gamma(t)=\begin{cases}
          -M\theta(\gamma(t)-(\frac{A^2}{2\theta})^{M/2}(M-1)!!) + (\frac{A^2}{2\theta})^{M/2} \sqrt{(2M-1)!! - ((M-1)!!)^2}\tilde{\xi}(t) \quad &\text{if} \, M \in 2\mathbb{Z} \\
          -M\theta \gamma(t) +  (\frac{A^2}{2\theta})^{M/2}\sqrt{(2M-1)!!}\tilde{\xi}(t)\quad &\text{if} \, M \in 2\mathbb{Z} + 1
 \\
     \end{cases}
\end{equation*}
\subsection{Correlations of the Wiener Process}
We wish to find the correlation between $\tilde{\xi}_\alpha(t)$, $\tilde{\xi}_\beta(t)$ such that $\left \langle \gamma_\alpha (t) \gamma_\beta(t)\right \rangle=\left \langle \eta^M_\alpha(t) \eta^M_\beta(t)\right \rangle$. We denote as $\rho = \text{Corr}(\xi_\alpha, \xi_\beta)=\text{Corr}(\eta_\alpha, \eta_\beta)$. We seek $c = \text{Corr}(\tilde{\xi}_\alpha, \tilde{\xi}_\beta)=\text{Corr}(\gamma_\alpha, \gamma_\beta)$. From the definition of correlation, and the moments of $\gamma_\alpha$ expressed above, we may say that 
\begin{align*}
    c = \frac{\left \langle \gamma_\alpha \gamma_\beta \right \rangle - \left \langle \gamma_\alpha \right \rangle \left \langle \gamma_\beta\right \rangle}{\sqrt{\text{var}(\gamma_\alpha)\text{var}(\gamma_\beta)}} = \frac{\left \langle \gamma_\alpha \gamma_\beta \right \rangle -  \bar{\gamma}_\alpha \bar{\gamma}_\beta}{\sqrt{((\frac{A_\alpha^2}{2\theta_\alpha})^M (2M-1)!! - \bar{\gamma}_\alpha^2)((\frac{A_\beta^2}{2\theta_\beta})^M (2M-1)!! - \bar{\gamma}_\beta^2)}} 
\end{align*}
The crossed second-moment is one of our constraints: we demand that
\begin{align*}
    \left \langle \gamma_\alpha \gamma_\beta\right \rangle = \left \langle \eta_\alpha^M \eta_\beta^M\right \rangle 
\end{align*}
Each $\eta$ is a Gaussian random variable of zero mean; we invoke Isserlis' theorem\cite{Isserlis1918}
\begin{align*}
    \left \langle \gamma_\alpha \gamma_\beta\right \rangle &= \left \langle \eta_\alpha^M \eta_\beta^M\right \rangle = \sum_{k=0}^M C_M(k) \left \langle \eta_\alpha \eta_\beta\right \rangle^k \left \langle \eta_\alpha^2\right \rangle^{\frac{M-k}{2}}\left \langle \eta_\beta^2\right \rangle^{\frac{M-k}{2}}\\
    \left \langle \gamma_\alpha \gamma_\beta\right \rangle &= \sum_{k=0}^M C_M(k) \rho^k \left \langle \eta_\alpha^2\right \rangle^{\frac{M}{2}-k}\left \langle \eta_\beta^2\right \rangle^{\frac{M}{2}-k}
\end{align*}
Where $C_M(k)$ is the number of ways for a partition of two groups of $M$ objects each, into pairs, to have $k$ "mixed" pairs (ie. one from each group):
\begin{align*}
    C_M(k) = [\binom{M}{k}(M-k-1)!!]^2 k!
\end{align*}
so long as $M-k$ is even. If it is odd, then the coefficient will be zero. We take the double factorial of a negative number to be equal to one. Thus, Wiener processes $\tilde{\xi}$ should have correlation
\begin{align*}
    c =  \frac{\sum_{k=0}^M C_M(k) \rho^k \left \langle \eta_\alpha^2\right \rangle^{\frac{M}{2}-k}\left \langle \eta_\beta^2\right \rangle^{\frac{M}{2}-k} -  \bar{\gamma}_\alpha \bar{\gamma}_\beta}{\sqrt{((\frac{A_\alpha^2}{2\theta_\alpha})^M (2M-1)!! - \bar{\gamma}_\alpha^2)((\frac{A_\beta^2}{2\theta_\beta})^M (2M-1)!! - \bar{\gamma}_\beta^2)}} 
\end{align*}
The only further note to add is that when considering covariances between even integer powers of Gaussian noise, recall that the noise terms are the fluctuations around mean, which has been subtracted off and incorporated into the deterministic part of the equation. That is, if we consider $\alpha(t), \beta(t)$, which we are using as substitutes for $\eta^M(t), \eta^P(t)$, where $M, P$ are even, then
\begin{equation*}
        \left \langle \alpha(t) \beta(t')\right \rangle = \left \langle \eta^M(t) \eta^P(t')\right \rangle - \left \langle \eta^M\right \rangle \left \langle \eta^P\right \rangle 
\end{equation*}
It may be necessary for the replacement noise to be a composite of OU processes, in order to maintain correlation structure with all other species of noise. 

\section{Lognormal moment closure} \label{lognormal}
We would like to make a brief note on the generic moments of random variables drawn from jointly-lognormal distributions, as this simple derivation, to our knowledge, is not explicitly stated anywhere, and may be useful to the reader.
Let us begin by saying that $X$ and $Y$ are jointly log-normal random variables. This means that $X = e^U$, $Y=e^V$, where $U$ and $V$ are normal random variables. If $X=e^U$ is log-normal, then $X^k = e^{kU}$ is also log-normal for some (not necessarily integer) number $k$, because $kU$ will be normally distributed. We know, by the properties of the moments of log-normal variables \cite{Halliwell2015}, that
\begin{align*}
    \left \langle X^k\right \rangle = e^{k\mu_X + k^2 \frac{\sigma_X^2}{2}}
\end{align*}
We wish to write the $k$th moment in terms of the first two moments; 
\begin{align*}
    \left \langle X\right \rangle &= e^{\mu_X + \frac{\sigma_X^2}{2}}\\
    \left \langle X^2\right \rangle &= e^{2 \mu_X + \frac{4 \sigma_X^2}{2}}
\end{align*}
Where $\mu_X$ and $\sigma_X^2$ are the mean and standard deviation of the underlying Gaussian $U$. This is sufficient to write those moments of $U$ in terms of moments of $X$:
\begin{align*}
    \mu_X &= \frac{1}{2}\ln(\frac{\left \langle X\right \rangle^4}{\left \langle X^2\right \rangle})\\
    \sigma_X^2 &= \ln(\frac{\left \langle X^2\right \rangle}{\left \langle X\right \rangle^2})\\
    \varrho = \frac{1}{\sigma_X \sigma_Y}\frac{\left \langle XY\right \rangle}{\left \langle X\right \rangle \left \langle Y\right \rangle} &= \frac{1}{\sqrt{ \ln(\frac{\left \langle X^2\right \rangle}{\left \langle X\right \rangle^2}) \ln(\frac{\left \langle Y^2\right \rangle}{\left \langle Y\right \rangle^2})}}\frac{\left \langle XY\right \rangle}{\left \langle X\right \rangle \left \langle Y\right \rangle}
\end{align*}
Where $\varrho$ is some possible correlation between the underlying $U$ and $V$.
In the generic case:
\begin{align*}
    \left \langle X^l Y^k\right \rangle = \left \langle X\right \rangle^{2l-l^2 -lk} \left \langle X^2\right \rangle^{l(l-1)/2}\left \langle Y\right \rangle^{2k-k^2 -lk} \left \langle Y^2\right \rangle^{k(k-1)/2}\left \langle XY\right \rangle^{kl}
\end{align*}

% ------------------------------------------------------------------------------
% Reference and Cited Works
% ------------------------------------------------------------------------------
\bibliography{references}
% ------------------------------------------------------------------------------
\end{document}